\documentclass[letterpaper]{article} 

\usepackage[usenames,dvipsnames]{color}
\usepackage{bm}
\usepackage{graphicx,fancyhdr,lscape,dsfont,amsmath,amssymb} 
\usepackage{color, amsthm}
\usepackage{longtable}
\usepackage{psfrag}
\usepackage{multirow}
\usepackage{stmaryrd}
\usepackage{appendix}
\usepackage{ulem} 
\usepackage{caption}
\usepackage{subcaption}
\usepackage{graphics,graphicx}
\usepackage{floatrow}

\allowdisplaybreaks

\newcommand{\vect}[1]{\vec{#1}}
\newcommand{\tensor}[1]{  {\bold {#1}} } 

\newcommand{\determinant}[1]{ {\rm det} \left[ \, {#1} \, \right] }

\newcommand{\totaltimed}[1]{ {{\rm d} \, {#1} \over {\rm d} t}   }

\newcommand{\Id}{ \tensor{1} }
\newcommand{\identity}{\mathds{1}}

\newcommand{\Def}{ \tensor{F} }

\newcommand{\firstpiola}{ \tensor{P} }
\newcommand{\secondpiola}{ \tensor{S} }

\newcommand{\rightCG}{ \tensor{C} }

\newcommand{\massflux}{ \vect{h} }
\newcommand{\heatflux}{ \vect{q} }
\newcommand{\heatsource}{ s_{q} }
\newcommand{\newheatfluxR}{\mbox{${\vect{q} {\mskip-12mu  - \mskip-3mu {{_R}} }} \,$}}

\newcommand{\entropy}{ \eta }

\newcommand{\helmholtz}{ \psi }

\newcommand{\thermalexp}{ \gamma }

\newcommand{\uchempot}{\mbox{$ {^u}{\mskip-2mu}\mu$}}
\newcommand{\etachempot}{\mbox{$ {^\eta}{\mskip-2mu}\mu$}}

\newcommand{\mobility}{\mbox{${\rm u} \mskip-8mu  | \,$}}

\newcommand{\myK}{\mbox{$\bold{K} \mskip-10mu  | $ \,}}

\newcommand{\Vint}{\mbox{${\sf V} $}}

\newcommand{\Divergence}[1]{ {\rm Div} \left[ \, {#1} \, \right] }
\newcommand{\Gradient}[1]{  {\rm Grad }\left[ \, {#1} \, \right] }

\newcommand{\Traction}{{\vect{t}_R}}

    \graphicspath{{./}{./Figures/}} 

\begin{document}

\title{On the chemo-thermo-mechanics of constrained reactive mixtures of solids. }
\author{ Alberto Salvadori$^{1,2}$, Mattia Serpelloni$^{1,2}$, 
 Robert M. McMeeking$^{1,3,4}$ 
\\
\begin{small}
$^{1}$The Mechanobiology Research Center, UNIBS - Brescia, 25123, Italy
 \end{small}
\\
\begin{small}
$^{2}$Department of Mechanical and Industrial Engineering,
 \end{small}
 \\
 \begin{small}
Universit{\`a} degli Studi di Brescia, via Branze 38, 25123 Brescia, Italy
 \end{small}
\\
 \begin{small}
$^{3}$Materials and Mechanical Engineering Departments, University of California, Santa Barbara, CA 93106, USA
 \end{small}
 \\
\begin{small}
$^{4}$School of Engineering, University of Aberdeen, King's College, Aberdeen AB24 3UE, UK
\end{small}
\\
}

\maketitle


\begin{abstract}

Building upon the classical chemo-mechanical theory of Larch{\'e} and Cahn for equilibrium, numerous studies have investigated the transport of species in solids, with or without trapping phenomena. In most applications --- such as the swelling of hydrogels, hydrogen embrittlement in metals, and the transport of lithium or sodium in battery electrodes --- the formation of a new phase or compound can be directly associated with the concentration of the diffusing species. 
In the present work, we focus on the formation of solid mixtures made of multiple compounds, each characterized by its own volumetric expansion coefficient. Such a scenario arises, for instance, during the sodiation of tin anodes, among other systems. The classical chemo-mechanical framework is naturally recovered as a particular case of the proposed formulation.

The theoretical framework developed herein elucidates and differentiates the concepts of phases and flowing species, while establishing rigorous connections between them. The present note is restricted to the general formulation of the governing equations, whereas  application-specific developments will be addressed in forthcoming publications.

\end{abstract}

%
%
%

\section{Introduction}
\label{sec:intro}


Over the past few decades, a broad spectrum of scientific and engineering applications has 
emerged in which tightly coupled physico-chemical processes --- such as mass transport, 
mechanical deformation, chemical reactions, thermal decomposition, phase transformation, and 
gas-generation reactions --- govern the behavior of solid mixtures. These include, among many others, 
mechanobiological phenomena such as cellular motility \cite{Hohmann2019,Ballestrem:2000aa} during 
angiogenesis \cite{BarrasaJRI2022,CarmelietJain} and metastasis \cite{StueltenEtAlNATURE2018,TOSIN2010969}, 
embryonic development \cite{Arroyo2019,MARSTON2006392}, biofilm formation 
\cite{CHOCKALINGAM2024105627}, and blood clotting \cite{Gupta:2023aa,Kim:2017aa,AlberJRInt2008}. 
Additional examples arise in metallurgy, such as carburizing and nitriding processes 
\cite{CAVALIERE200926,Woowiec-KoreckaAMSE2023,YangNIT2012}; in energy storage, particularly 
in the operation of rechargeable batteries 
\cite{SalvadoriEtAlJMPS2013,AFSHAR2021104368,LatzZauschBJN2015}; and in the combustion of 
solid rocket propellants \cite{SuttonRocketPropulsion,KubotaPropellants,DENNIS2019115646}. 
These heterogeneous applications illustrate the pervasive importance of reactive solid mixtures 
across disciplines.

Mixture theory is a powerful framework for addressing complex systems with multiple interacting continua. Foundational contributions can be attributed to Truesdell \cite{truesdell1957sulle,truesdell1957sulleII}, Bowen \cite{bowen1976theory,bowen1979theory,bowen1980incompressible,bowen1982compressible}, Green and Naghdi \cite{green1965dynamical,green1969basic}, as well as Rajagopal and Tao \cite{rajagopal1995mechanics}. Truesdell and Toupin \cite{TruesdellToupin1960} rigorously formulated conservation laws for chemically reactive heterogeneous mixtures, with extensions reviewed by Bedford and Drumheller \cite{bedford1983theories}.  According to \cite{malek2020theory}, two main frameworks within the theory of interacting continua can be identified: (i) {\it{Continuum Mixture Theory}} and (ii) {\it{Multi-phase Theory}}. The former is based on the {principle of equi-presence}, assuming that all components coexist at every point in the continuum. Multi-phase Theory, in contrast, relies on single-component balance laws combined with interface conditions at fine spatial scales, such as the pore-scale in porous media, where constituents are distinguishable. Quantities of interest are then upscaled via averaging operators, to form a macroscopic continuum mixture model \cite{drew2006theory}.
Fundamental references on averaging methods include \cite{bowen1976theory,BOWEN1982697,BEDFORD1983863,wang2012rational,bear1990introduction,morland1992flow,de1996highlights}, while macroscale theories trace back to Biot's pioneering work \cite{biot1941general,biot1956theory,biot1972theory,biot1977variational} and its extensions by \cite{coussy1995mechanics,dormieux1995macroscopic}. 
Some authors \cite{Gouin} raised fundamental questions about entropy, thermodynamic laws, and temperature definitions in mixtures, topics extensively discussed by \cite{bowen1979theory,muller1967entropy,williams1973theory,atkin1976continuum,sampaio1977viscosities,nunziato1980ideal,drew1982mathematical}.

Following \cite{malek2020theory,hutter2013continuum}, mixtures can be categorized by the complexity of their balance laws. Four classes are identified: {Class I} involves $N$ mass balances, a single momentum balance, and one energy and one entropy balance for the mixture (examples: Fickian diffusion, dilute gases, pollution). {Class II} considers $N$ mass and momentum balances but only one energy and entropy balance (examples: porous media flow, chemical reactors). {Class III} includes $N$ balances of mass, momentum, and energy, but one entropy balance (examples: plasmas, nonequilibrium melts). {Class IV} extends entropy balances to each constituent individually, though this is rarely applied, with most models using a mixture-wide second law.

In their pioneering work  \cite{LarcheCahn1973} Larch{\'e} and Cahn developed a Continuum Mixture Theory of Class I for the thermodynamic equilibrium of a multi-component solid. In \cite{LarcheCahn1978} they
coupled the flow of species in a hosting solid material to its finite strains geometrical evolution. Their approach, which accounts for large volumetric deformations and mass balance, is a milestone in the chemo-mechanics of materials \cite{ARRICCA2023105425,SalvadoriEtAlJMPS2018} and paved the way toward the formulation of homogenized, single phase continuum theories out of thermodynamic equilibrium. Such theories have been applied to numerous phenomena in mechanobiology, such as tumor growth and biofilms \cite{CHOCKALINGAM2024105627}, protein relocation \cite{membranes12070652,SerpelloniEtAlAMS2021,SerpelloniEtAlIJES2022}, actin-based motility \cite{SalvadoriSR2024,BONANNO2023105273},  as well as in other research fields such as hydrogen embrittlement \cite{Anand2011,DiLeoAnand2013}, hydrogels \cite{HamzaviDrozdovGuBirgersson}, and energy storage materials \cite{Drozdov2014b,AFSHAR2021104368,Ganser2019}. In all those works, Larch{\'e} and Cahn theory has been enriched with multiplicative decomposition of the strain tensor \cite{AMBROSI20021297,ChesterAnandJMPS2010}, to account for the volumetric swelling associated with mass transport and/or chemical reactions. We too exploit multiplicative decomposition in our theory, deriving a rationale for the evaluation of the change in volume due to evolution of species concentration.


Whereas in \cite{ateshian2007theory} the term \textit{reactive} denotes species that participate in chemical reactions at phase interfaces, in the present work it is used instead to describe the emergence of new solid phases produced by reactive species within the bulk. To capture such phenomena, we develop a finite-strain, chemo-mechanical continuum formulation for the time-dependent (i.e., out-of-equilibrium) evolution of solid mixtures, built on the fundamental assumption that all solid phases share a common velocity field and therefore exhibit no relative motion. 
This viewpoint distinguishes our framework from unconstrained mixture theories, where each constituent may move independently \cite{truesdell1957sulle,truesdell1957sulleII,TruesdellToupin1960}, as well as from constrained mixtures in which the constituents move together only in the current configuration \cite{humphrey2002constrained,Humphrey:2021aa}. By postulating a single motion governing the kinematics of the entire body, a unique reference configuration naturally follows. As discussed in Section~\ref{subsec:LC}, the classical network theory of Larch\'{e} and Cahn --- originally formulated for equilibrium chemo-mechanical processes \cite{LarcheCahn1978,LarcheCahn1973} --- together with subsequent extensions incorporating transport and chemical reactions \cite{ChesterEtAlIJSS2015,ChesterAnandJMPS2010,AnandJMPS2012,Drozdov2014b,AFSHAR2021104368,SalvadoriEtAlJMPS2018,ARRICCA2023105425}, can be recovered as a special case of the more general framework proposed here.

%

Section~\ref{sec:defass} introduces the kinematic assumptions underpinning our framework, including the classical multiplicative decomposition of the deformation gradient. This formulation naturally yields a swelling response that departs from standard treatments in the literature and represents one of the principal contributions of the present work, with direct implications for the resulting governing equations. These equations are derived in their Lagrangian form in Section~\ref{governing_equations}, based on the balance laws for reaction kinetics, mass, and momentum established in Section~\ref{sec:Balancelaws}, together with the thermodynamic analysis of transport, chemical reactions, and mechanical coupling developed in Section~\ref{sec:thermo}. Final remarks are provided in Section~\ref{sec:conclisions}.

\section{Definitions and kinematic assumptions}
\label{sec:defass}

\subsection{Phases, species, configurations, volume fractions.}
\label{subsec:ref_curr}

In classical thermodynamics, the notions of {\it{phase}} and {\it{species}} are fundamental: a phase denotes a physically homogeneous and mechanically separable portion of matter, whereas a species identifies a chemically distinct constituent whose amount may vary within or across phases. These concepts follow the standard treatments in \cite{CallenTD,deGrootBook,ShellBook2015}, and the present note does not seek to introduce new definitions for such foundational ideas.
As an illustrative example, we may consider the microstructural evolution of tin nanoparticles during electrochemical sodiation\footnote{We term sodiation the transport of atomic sodium in the anode. In fact, we may assume that transport of ${\rm NaSn_x} $ alloys via solid-state diffusion is a secondary effect, for the steric dimension of those large molecules. }, as investigated by \cite{WangEtAlNANOLETTERS2012} using \textit{in situ} transmission electron microscopy. The study revealed that the initial sodiation proceeds in two distinct stages. In the first stage, crystalline Sn nanoparticles undergo a two-phase reaction with a migrating phase boundary, forming a Na-poor amorphous $\mathrm{Na_xSn}$ alloy ($x \approx 0.5$). In the subsequent stage, this alloy is further sodiated into several Na-rich amorphous phases, ultimately crystallizing into $\mathrm{Na_{15}Sn_4}$ through a single-phase mechanism. The corresponding volumetric expansion was approximately $60\%$ during the first stage and increased to about $420\%$ after the second.



\begin{figure}[!htb]
\centering
\includegraphics[width=0.75\linewidth]{FmFc.pdf}
\caption{ A schematic of the multiplicative decomposition $\tensor{F} =   \tensor{F}^{m} \tensor{F}^{c}$ and related configurations.  }
\label{fig:mult_decomp}
\end{figure}

\bigskip
Define $V(t)$ as a region where a solid exists, which may result in different coexisting phases. 
As is customary \cite{ateshian2007theory}, we assume that no relative motion occurs among the solid phases, so that their individual velocities $\vect{v}^{\,\alpha}$ coincide at all times and are represented collectively by $\vect{v}$. Since a single motion characterizes the kinematics of the body, a single reference configuration for $V$ can be defined. It is illustrated in Fig.~\ref{fig:mult_decomp}. 
Let $V_R$ denote the referential image of the current configuration $V(t)$, such that for every spatial point $\vect{x} \in V(t)$ there exists a unique material point $\vect{X} \in V_R$ and an invertible, sufficiently smooth mapping $\vect{\chi}$ --- termed the \textit{motion} --- that satisfies
\[
\vect{x} = \vect{\chi}(\vect{X}, t).
\]
When attention is restricted to a fixed time $t$, the motion is termed {\it{deformation}} at time $t$ and denoted by $\vect{\chi}_t(\vect{X})$. 
As usual, the deformation gradient is defined as $\tensor{F} = \partial \vect{\chi}_t / \partial \vect{X}$, with $J = \determinant{\tensor{F}} > 0$. Henceforth, the subscript $_R$ denotes any field expressed in the reference configuration, corresponding to its spatial counterpart.


\bigskip
%
When multiple phases coexist in $V(t)$, the material points $\vect{X}$ and their images $\vect{x}$ may be interpreted through the lens of a Representative Volume Element (RVE) \cite{torquato2002}, thereby enabling one to regard the body as a {\it{solid mixture}}, that is, a collection of distinct phases \cite{CapaldiBook}. By the adopted definition of phases, the corresponding subdomains within the RVE are understood to be {\it{separable, independent, and uncompounded with each other}}\footnote{Quoted from \cite{COWIN201247}, p.~2}.
The notion of volume fraction, denoted by the symbol $\Phi^\alpha$, defines the amount of the volume occupied by each phase $\alpha$ and is defined as
\begin{equation}
\label{eq:def:volume_fractions}
\Phi^\alpha 
= \frac{{\rm d} V^\alpha }{{\rm d} V}
\; ,
\qquad
\Phi^\alpha_R 
= \frac{{\rm d} V_R^\alpha }{{\rm d} V_R}
\; .
\end{equation}
As will be discussed in a subsequent section, the referential and current volume fractions need not coincide, even though all phases undergo a single motion.
Furthermore,  ${\rm d} V = \sum_\alpha {\rm d} V^\alpha$ and ${\rm d} V_R = \sum_\alpha {\rm d} V^\alpha_R$. Hence
\begin{equation}
\label{eq:prop:volume_fractions}
 \sum_\alpha \Phi^\alpha
= 
 \sum_\alpha \Phi^\alpha_R 
 =
 1
\; .
\end{equation}

\bigskip
Solid phases may consist of {\it{species}} $\beta$, capable of moving within volumes $V$ and mixing in arbitrary proportions to form a phase. In the classical chemo-mechanical framework established by Larch{\'e} and Cahn for equilibrium \cite{LarcheCahn1978,LarcheCahn1973}, and in its subsequent extensions addressing transport and chemical reactions \cite{ChesterEtAlIJSS2015,ChesterAnandJMPS2010,AnandJMPS2012,Drozdov2014b,AFSHAR2021104368,SalvadoriEtAlJMPS2018,ARRICCA2023105425}, a single species is typically assumed to migrate within a pristine and homogenous pre-existing lattice. This description applies, for example, to hydrogen diffusion in metals \cite{DiLeoAnand2013}, as well as to lithiation \cite{Ganser2019} or sodiation processes occurring in the active particles of intercalation-type electrodes in electrochemical batteries. 
In conversion-type electrodes\footnote{Conversion-type electrodes are a class of materials that operate through chemical conversion reactions with ions, rather than through the reversible insertion or extraction mechanisms characteristic of intercalation-type electrodes.}, the distinction between the pristine active material and its lithiated or sodiated compounds has, in some cases \cite{AFSHAR2021104368}, been directly associated with the local concentration of lithium or sodium --- thus effectively identifying the phases with the species themselves. In this work, by contrast, we draw a clear distinction between the various lithiated or sodiated compounds (the \textit{phases}) and the \textit{species}\footnote{This distinction evokes a multiscale perspective, wherein phases correspond to the macroscopic, averaged response of the system to external actions and boundary conditions, while species pertain to the microscopic level where molecular interactions occur. In this note, however, we do not address scale transitions or homogenization procedures as in \cite{MillerPenta2020,10.1121/1.386945}; instead, we confine our attention to a single scale.}, namely lithium or sodium, which migrate within each phase.

\bigskip 
Changes in the concentration of reactive species in $V_R$ arise from referential mass transport and chemical reactions. Once reactions occur, the resulting concentration variations induce corresponding changes in the volume fractions of the phases. 
It is well established in the chemo-mechanical literature that such processes do not alter the referential volume, even though they modify the current volume $V(t)$ (see, for instance, \cite{BONANNO2023105273,ChesterEtAlIJSS2015,dileoetalJMPS2014}, among many others). This convention preserves a \textit{stress-free} reference configuration during chemical reactions and is equivalent to assuming that the referential molar volumes of all phases are arbitrary yet equal. 
A key implication of this assumption is that the reference configuration becomes an abstract, idealized construct, rather than the actual physical configuration at the initial time. The present theory is founded upon this premise, too.


\bigskip
Denote by $\Omega^\beta$ the molar volume of the species $\beta$ and by $N^\beta$ the number of moles of the same species. Since the referential molar volumes of all species are arbitrary and equal, $  {\Omega}^\beta_R = {\overline{\Omega}}_R$. 
The molar concentration\footnote{Multiplication of the concentration by the molar mass provides the {\it{apparent density}}, borrowing the terminology from \cite{ateshian2007theory}. } for the species $\beta$ is defined as the number of moles per unit {\it{total}} volume either in the current or in the reference configuration, i.e.,
\begin{equation}
\label{eq:def:concentration}
c^\beta = \frac{{\rm d}N^\beta }{{\rm d} V}
\; ,
\qquad
c^\beta_R = \frac{{\rm d}N^\beta_R }{{\rm d} V_R}
\; .
\end{equation}
In several circumstances, such as the mass action law, we shall refer to the concentration of a phase $\alpha$, which is defined as
\begin{equation}
\label{eq:def:phase_concentration}
 c_R^\alpha 
=
  \;  
 \frac{ \Phi^\alpha_R }{  {\overline{\Omega}}_R}
\end{equation}
in the reference configuration.

\bigskip
In this note, as in \cite{COWIN201247}, the mixture is considered to be an
open system. This means that  {\it{species}} are allowed to escape or enter any subpart ${\cal{S}}(t) \subset V(t)$. 
At a given time $t$ the deformation map $\vect{\chi}_t$ preserves the species content in the referential ${\cal{S}}_R$ and current  ${\cal{S}}(t) $ subparts,
i.e. 
$$
{\rm d}N_R^\beta
=
{\rm d}N^\beta
\; .
$$
As a consequence,
\begin{equation}
\label{eq:prop:concentrations}
c^\beta 
= \frac{{\rm d}N^\beta }{{\rm d} V} 
= \frac{{\rm d}N^\beta }{J {\rm d} V_R} 
= \frac{{\rm d}N^\beta_R }{J {\rm d} V_R} 
= \frac{c^\beta_R  }{J  }
\; .
\end{equation}
Since eq. \eqref{eq:prop:concentrations} holds for any ${\rm d}N^\beta$, it applies also to the maximum number of moles that can be hosted in the solid mixture. Accordingly, we shall conclude that
\begin{equation}
\label{eq:prop:cmax}
c^\beta_{sat} 
= \frac{c^\beta_{sat,R}  }{J  }
\; .
\end{equation}

\bigskip
Since 
\begin{equation}
\label{eq:def:ref_volume_fraction}
{\rm d} V_R^\beta = 
{\rm d}N_R^\beta  \,  {\overline{\Omega}}_R =
c_R^\beta  \,  {\overline{\Omega}}_R \, {\rm d} V_R
\; ,
\end{equation}
species concentrations are not independent and it holds
\begin{equation}
\label{eq:def:ref_concentration_dependence}
 \sum_\beta c_R^\beta  
= 
{  {\overline{\Omega}}_R }^{-1}
\; .
\end{equation}
The inverse of the molar volume ${\overline{\Omega}}_R$  is thus an upper bound for the concentration of each species $\beta$. Because
the saturation concentration $c^\beta_{sat,R} $ for the same species can differ from, although not exceed, ${  {\overline{\Omega}}_R }^{-1}$, 
\begin{equation}
\label{eq:propr:how_cmax_changes}
c^\beta_{sat,R} 
=
{  {\overline{\Omega}}_R }^{-1}
\end{equation}
appears to be a natural and convenient definition. 
Note that $c^\beta_{sat,R} $ is then the same for all species and can be reached only in an homogeneous system made of the species $\beta$.
%
Provided that $ {\overline{\Omega}}_R$, which is arbitrary, is given unit value, $c^\beta_{sat,R} =1$.


\subsection{Intermediate configuration}
\label{subsec:intermediate}

\subsubsection{Swelling factor}

%
Mass transport and chemical reactions induce volumetric swelling and shrinking, which are generally captured by means of a multiplicative decomposition \cite{AMBROSI20021297,ChesterAnandJMPS2010}
\begin{equation}
\label{eq:multiplicative_decomposition}
	\tensor{F} =   \tensor{F}^{m} \tensor{F}^{c}
	\; .
\end{equation}
The {\it{swelling tensor}} $\tensor{F}^c$ is the local distortion of the neighborhood at point $\vect{X}$ after a phase change occurs \cite{SerpelloniEtAlAMS2021}. The mechanical\footnote{Not necessarily elastic. Further tensorial splittings can occur for $ \tensor{F}^{m}$ and will be discussed later in this paper.} tensor $ \tensor{F}^{m}$ recovers the compatibility of the network. As stated in \cite{GurtinFriedAnand} for the Kr{\"o}ner decomposition, the elastic and the swelling tensors are not gradients of point-field maps. In fact, a complete, intermediate and stress-free configuration in which referential vectors are mapped via $\tensor{F}^c$ is, in general, idealized, since, usually, it lacks geometrical compatibility.


\bigskip
To elucidate the kinematics of the solid mixture, we begin by examining chemical reactions and compound formation processes that occur in the absence of mechanical constraints, i.e., when $\tensor{F}^{m} = \identity$. In this case, the current configuration is stress-free and coincides with the intermediate configuration.
The volumetric change between reference (${\rm d} V^\alpha_R$) and intermediate (${\rm d} {\Vint}^\alpha$) configuration can be written in terms of the constant $\Omega^\alpha$, the molar volume of phase $\alpha$ in the absence of stress
\begin{equation}
\label{eq:prop:volume_per phase}
{\rm d} \Vint^\alpha
=
{\rm d}N^\alpha  \, \Omega^\alpha
=
{\rm d}N_R^\alpha  \; {\overline{\Omega}}_R \;  \frac{ \Omega^\alpha }{ {\overline{\Omega}}_R }
=
{\rm d} V^\alpha_R  \;  \omega^\alpha 
\end{equation}
where 
\begin{equation}
\label{eq:omega_alpha}
\omega^\alpha = \frac{ \Omega^\alpha }{  {\overline{\Omega}}_R }
\end{equation}
is, by  convention, a swelling factor for each phase\footnote{``By convention" means here that ${\overline{\Omega}}_R$ is uncorrelated to the real molar volumes. }. 
By summing Eq. \eqref{eq:prop:volume_per phase} on all phases in the mixture one obtains
\begin{equation}
\label{eq:prop:volume_swelling}
{\rm d} \Vint
=
\sum_\alpha {\rm d} \Vint^\alpha
=
\sum_\alpha {\rm d} V^\alpha_R  \;  \omega^\alpha 
=
 {\rm d} V_R \; \sum_\alpha \Phi^\alpha_R  \;   \omega^\alpha  
=
 {\rm d} V_R \; \sum_\alpha  c_R^\alpha \;  \Omega^\alpha
 \; .
\end{equation}
Equation \eqref{eq:prop:volume_swelling} highlights the observation that the determinant $J^c =  \determinant{\tensor{F}^c}$ amounts to
\begin{equation}
\label{eq:def:Jc}
J^c 
=
{\rm d} \Vint /  {\rm d} V_R 
=
 \sum_\alpha \Phi^\alpha_R  \;  \omega^\alpha
=
 \sum_\alpha  c_R^\alpha \;  \Omega^\alpha
 \; .
\end{equation}
%
Note that if the reference configuration is not identified with a real initial configuration, $J^c$ may differ from unity at $t=0$. This notion relates to the arbitrariness of ${\overline{\Omega}}_R$. The definition of $J^c$ introduced herein constitutes a principal result of this study, extending the classical expressions found in several prior works (e.g., \cite{ChesterAnandJMPS2010,Drozdov2014b,DiLeoAnand2013,ARRICCA2023105425,Ganser2019,AnandJMPS2012}).

\bigskip

The mechanobiological onset and progression of pathological conditions provide several opportunities for applying the theory developed thus far. Osteoarthritis, for example, is a pervasive and debilitating disease in which cartilage degeneration plays a central role. Its evolution can be described within a chemo--mechano--biological framework, as proposed in \cite{RAHMAN2023107419}. 
Cartilage is a heterogeneous solid, whose extracellular matrix (ECM) comprises a negatively charged proteoglycan network and a predominantly type~II collagen fiber architecture. Both constituents contribute to the tissue's mechanical stiffness and regulate fluid permeation. Their volume fractions evolve over time, particularly when homeostasis --- normally sustained by the mechanical stimuli of daily activities --- is disrupted, thereby inducing an imbalance between ECM synthesis and degradation.
Mechanical cues modulate chondrocyte biosynthetic activity in distinct ways: elevated cyclic hydrostatic pressure enhances proteoglycan production, whereas cyclic tensile strain promotes collagen synthesis. Conversely, insufficient mechanical loading suppresses chondrocyte activity --- prolonged immobilization, for instance, leads to ECM depletion and cartilage thinning --- while excessive loading similarly results in matrix degradation.
The evolution of the functional collagen volume fraction should comply with eq.~\eqref{eq:def:Jc}, although alternative formulations have been proposed\footnote{See in particular equation 4 developed therein.} in \cite{RAHMAN2023107419}. The systematic procedure introduced in this note may thereby provide a rigorous foundation for microstructurally informed continuum models of cartilage \cite{Pierce:2016aa,FEDERICO20083309,FedericoJRSInt2010}.

\bigskip
%
Metallurgy also provides significant applications for eq.~\eqref{eq:def:Jc}.  
The diffusion of carbon in steels, for example, plays a pivotal role in enhancing surface strength through martensite formation during quenching \cite{CAVALIERE200926}. 
{\itshape Carburizing} \cite{Woowiec-KoreckaAMSE2023} is a thermo–chemical treatment in which carbon diffuses into the iron lattice, promoting martensitic transformation in the outer layers during subsequent temper quenching. The formation of these hardened layers improves strength and fatigue resistance, owing both to increased hardness and to the development of compressive residual stresses on the surface. The thermo–chemical stage of the treatment consists of high–temperature carbon diffusion at the surface, typically decomposed into (i) the reaction of carbon with the initially low–concentration surface and (ii) its subsequent diffusion into the bulk of the component.
{\itshape Nitriding} \cite{YangNIT2012} is another thermo–chemical surface treatment in which nitrogen is introduced from an ammonia atmosphere into steel at temperatures below the eutectoid point. Because nitrogen has low solubility in ferrite, nitrides precipitate during the process. As a result, a compound layer and an underlying diffusion zone form near the surface. The compound layer substantially improves wear and corrosion resistance, as well as fatigue endurance.

\subsubsection{Transformations of concentrations and volume fractions.}
\label{subsubsec:transformation}

Concentrations transform between reference and intermediate configurations via $J^c$. It holds, in fact, 
\begin{equation}
\label{eq:prop:int:concentrations}
c^\beta_{\Vint}
= \frac{{\rm d}N^\beta }{{\rm d} \Vint} 
= \frac{{\rm d}N^\beta }{J^c \, {\rm d} V_R} 
= \frac{{\rm d}N^\beta_R }{J^c \, {\rm d} V_R} 
= \frac{c^\beta_R  }{J^c }
\; .
\end{equation}
The path of reasoning taken in equations \eqref{eq:prop:concentrations} and \eqref{eq:prop:cmax} naturally leads to define 
\begin{equation}
\label{eq:intermediate:saturation:2}
c^\beta_{sat,\Vint}
= \frac{c^\beta_{sat,R}}{J^c}
= \frac{1}{\overline{\Omega}_R\, J^c}
\; ,
\end{equation}

We shall point out that the definition of saturation concentration $c^\beta_{sat,\Vint}$ as the push-forward of the referential value $c^\beta_{sat,R}$ shall be understood as an ``apparent'', or  ``mixture-based’'' saturation concentration. Equation \eqref{eq:intermediate:saturation:2} provides a saturation concentration which {\it{is not}} an intrinsic
property of species $\beta$, being affected by the presence of other
constituents in the mixture through $J^c$ --- see eq. \eqref{eq:def:Jc}.  

In fact, since the saturation concentration defined in
eq.~\eqref{eq:propr:how_cmax_changes} can be attained only in a
homogeneous system --- namely, when $J^c = \omega^\beta$ --- the ``real"
saturation concentration of species $\beta$ in the intermediate
configuration reads
\begin{equation}
\label{eq:intermediate:saturation}
c^{\beta,real}_{sat,\Vint}
= 
\frac{1}{\omega^\beta\,\overline{\Omega}_R}
= \frac{1}{\Omega^\beta}
\; .
\end{equation}
Equation \eqref{eq:intermediate:saturation} provides a saturation concentration as an intrinsic
property of species $\beta$, unaffected by the presence of other
constituents in the mixture.  

%

\bigskip
We shall highlight here that since a single motion characterizes the kinematics of the body, $J^c$ applies to a whole neighborhood but it does not apply separately to each phase $\alpha$: 
$$
{\rm d} \Vint^\alpha =   \omega^\alpha  \; {\rm d} V^\alpha_R  
\ne
J^c \, {\rm d} V^\alpha_R 
\; .
$$
Accordingly, volume fractions do not transform via $J^c$ as concentrations.
Referential and intermediate volume fractions are rather related by
\begin{equation}
\label{eq:propr:ref_volume_fraction}
\Phi^\alpha_{\Vint} 
= \frac{{\rm d} \Vint^\alpha }{{\rm d} \Vint}
= \frac{  \omega^\alpha  \; {\rm d} V^\alpha_R  }{ J^c \, {\rm d} V_R}
= \frac{  \omega^\alpha  }{ J^c } \; \Phi^\alpha_R \; 
\qquad
\Rightarrow
\qquad
\Phi^\alpha_R 
= \Phi^\alpha_{\Vint} \; \frac{ J^c }{  \omega^\alpha  }
\; .
\end{equation}
Consistency is granted since in view of eq. \eqref{eq:def:Jc} it holds
\begin{equation}
\label{eq:def:intermediate_volume_fraction_compatibility}
 \sum_\alpha \Phi^\alpha_{\Vint}  
 = \sum_\alpha   \frac{  \omega^\alpha  }{ J^c } \; \Phi^\alpha_R \; 
 =  \frac{  1  }{ J^c } \; \sum_\alpha \omega^\alpha  \; \Phi^\alpha_R \; 
= 1
\; .
\end{equation}



\subsection{Mechanical deformations}
\label{subsec:uniform}

As per the so-called Taylor or Voigt macro to micro scale transition in computational homogenization \cite{GEERSIJMC2003,GEERSJCAM2010,GeersEtalencyclopedia2015}, we assume that all (solid) phases experience the same mechanical deformation. In such a case, 
\begin{equation}
\label{eq:def:Jm}
J^m 
=  
\determinant{\tensor{F}^m}
=
 {\rm d} V  / {\rm d} \Vint 
=
 {\rm d} V^\alpha  / {\rm d} \Vint^\alpha 
\end{equation}
and
\begin{equation}
\label{eq:def:intermediate_volume_fraction}
 {\rm d} \Vint^\alpha   / {\rm d} \Vint 
=
 {\rm d} V^\alpha  /  {\rm d} V
 \; .
\end{equation}
%
In contrast to the general framework of mixture theory \cite{TruesdellToupin1960}, adopting a single motion to describe the kinematics of the body implies that {\it{the volume fraction of each phase in the intermediate configuration remains invariant under purely mechanical deformations}}.
Furthermore, in view of Eq. \eqref{eq:prop:volume_per phase}, it holds 
\begin{equation}
\label{eq:prop:volume_per_phase_with_mech}
{\rm d} V^\alpha
=
J^m \, {\rm d} \Vint^\alpha
=
 J^m \, \omega^\alpha \; {\rm d} V^\alpha_R  
\; .
\end{equation}
In view of Eq. \eqref{eq:prop:volume_per_phase_with_mech}, referential and current volume fractions are related by
\begin{equation}
\label{eq:propr:ref_volume_fraction:2}
\Phi^\alpha 
= \frac{{\rm d} V^\alpha }{{\rm d} V}
= \frac{ J^m \, {\rm d} \Vint^\alpha }{{\rm d} V}
= \frac{ J^m \, {\rm d} V^\alpha_R  \;  \omega^\alpha  }{ J \, {\rm d} V_R}
= \Phi^\alpha_R \; \frac{  \omega^\alpha  }{ J^c }
= \Phi^\alpha_{\Vint} 
\; .
\end{equation}
Differently from non-reactive mixtures, therefore, although the motion is unique {\it{the volume fraction of every phase in the reference configuration is not conserved in the current configuration}}, and more specifically is not conserved in the intermediate configuration.
Consistency is granted since
\begin{equation}
\label{eq:def:curr_volume_fraction_compatibility}
 \sum_\alpha  \Phi^\alpha 
 =  \sum_\alpha \Phi^\alpha_{\Vint}  
= 1
\; .
\end{equation}
We can estimate from eq. \eqref{eq:propr:ref_volume_fraction} the ratio of the volume fractions of two species ${\alpha_1}$ and ${\alpha_2}$  as
\begin{equation}
 \frac{  \Phi^{\alpha_1}_R   }{  \Phi^{\alpha_2}_R   }
 =
 \frac{  \Phi^{\alpha_1}  }{  \Phi^{\alpha_2}   } \; \frac{  \omega^{\alpha_2} }{  \omega^{\alpha_1}  }
 =
 \frac{  \Phi^{\alpha_1}  }{  \Phi^{\alpha_2}   } \; \frac{  \Omega^{\alpha_2} }{  \Omega^{\alpha_1}  }
 \; .
\end{equation}
If for instance two species occupy half of the reference configuration $\Phi^{\alpha_1}_R=\Phi^{\alpha_2}_R=0.5$ and the molar volume of ${\alpha_2}$ is as twice as the molar volume of ${\alpha_1}$, then  $\Phi^{\alpha_1}=0.5 \, \Phi^{\alpha_2} = 1/3$.
Similarly, when moles of a phase ${\alpha_1}$ are converted into moles of another phase ${\alpha_2}$, e.g.,  because of a chemical reaction rate $\dot{N}$, the volumes change as
\begin{equation}
\label{eq:prop:volume_swelling_per phase}
{\rm d} \Vint^{\alpha_1}
=
\dot{N} \; {\rm d}t  \; \Omega^{\alpha_1}
=
\dot{N} \; {\rm d}t  \; \Omega^{\alpha_2} \;  \frac{ \Omega^{\alpha_1} }{  \Omega^{\alpha_2}  }
=
{\rm d} \Vint^{\alpha_2}  \;  \frac{ \Omega^{\alpha_1} }{  \Omega^{\alpha_2}  }
\; .
\end{equation}
We will define $ \omega^{{\alpha_1}/{\alpha_2}} =  \frac{ \Omega^{\alpha_1} }{  \Omega^{\alpha_2}  } =  \frac{ \omega^{\alpha_1}  }{  \omega^{\alpha_2}  }$.

\bigskip
From equations \eqref{eq:prop:concentrations} and \eqref{eq:prop:int:concentrations}, one immediately finds that
\begin{equation}
\label{eq:prop:int:2:concentrations}
c^\beta 
= \frac{c^\beta_R  }{J  }
= \frac{c^\beta_{\Vint} }{J^m }
\; .
\end{equation}
According to this statement and following the discussion made in section \ref{subsubsec:transformation}, the {\it{saturation}} concentration of every species in a mixture in the current configuration is
\begin{equation}
\label{eq:def:cmax}
c_{sat}^\beta = 
\frac{c^\beta_{sat,\Vint}  }{ J^m  } 
\; ,
\qquad
c_{sat}^{mix} = \frac{c^{mix}_{sat,\Vint}   }{ J^m  } =  \frac{ c^\beta_{sat,R} }{ J  } 
\; .
\end{equation}
%

\subsection{The network model}
\label{subsec:LC}

The network 
model proposed by Larch{\'e} and Cahn 
and in its subsequent extensions addressing transport and chemical reactions 
can be recovered from this general framework. 
As an illustration, assume that the system is made of the pristine hosting material (denoted by 0) and the phase (denoted by 1) of hosting material filled by the transported specie. Such system has been used to model hydrogen diffusion in solids \cite{DiLeoAnand2013}, lithiation in battery electrodes \cite{Drozdov2014b}, and fluid permeation in elastomeric materials \cite{ChesterAnandJMPS2010}.
From eq. \eqref{eq:def:Jc}, 
\begin{equation}
\label{eq:intermezzo:Jc:2}
J^c 
=
 \Phi^0_R  \;  \omega^0 +  \Phi^1_R  \;  \omega^1
=
 (1-\Phi^1_R)  \;  \omega^0 +  \Phi^1_R  \;  \omega^1
=
  \omega^0 +  \Phi^1_R  \; ( \omega^1 -  \omega^0 )
 \; .
\end{equation}
Several authors (see  \cite{ARRICCA2023105425,Ganser2019,ChesterAnandJMPS2010,AnandJMPS2012} among may others) write eq. \eqref{eq:intermezzo:Jc:2} in terms of the concentration of the diffusing species rather than volume fraction of phases\footnote{ Such identification of species and phases may be misleading if the insertion and diffusion of species can lead to different intermediate compounds with different molar volumes, as occurs, e.g., for sodium in tin anodes \cite{WangEtAlNANOLETTERS2012}. }.
To this aim, assume that the whole body is made of pristine hosting material, i.e. $ \Phi^1_R=0$, at $t=0$ and that the reference and the initial configuration coincide. The requirement that $J=1$ at $t=0$ implies $ {\overline{\Omega}}_R=\Omega^0$ from definition \eqref{eq:omega_alpha}. Equation \eqref{eq:intermezzo:Jc:2}, making use of eq. \eqref{eq:def:phase_concentration}, thus is restated as
\begin{equation}
\label{eq:intermezzo:Jc:3}
J^c 
=
 1 +  \Phi^1_R  \; ( \omega^1 -  1 )
=
 1 +  c_R^1   \; ( \Omega^1 -  \Omega^0 )
 \; .
\end{equation}
The parameter
\begin{equation}
\label{eq:intermezzo:Jc}
J^c -1 
=  
\frac{{\rm d} \Vint -  {\rm d} V_R}{  {\rm d} V_R }  
=
 c_R^1   \; ( \Omega^1 -  \Omega^0 )
 \; ,
\end{equation}
measures the relative change in volume going from the reference to the intermediate configuration due to the concentration of species.

\section{Balance laws }
\label{sec:Balancelaws}

In the present framework, we retain the same assumptions adopted in \cite{SalvadoriEtAlJMPS2018,ARRICCA2023105425}. Species are allowed to diffuse freely through the host phases, with diffusivities that may differ from one phase to another and may even vanish, leading to species trapping. Species transport is therefore ascribed solely to motion through lattice (interstitial) sites. 
Accordingly, the flux of species $\alpha$ --- defined as the number of moles crossing a unit area per unit time and denoted by $\massflux_{\alpha}$ --- arises exclusively from interstitial lattice diffusion, in accordance with \cite{LarcheCahn1973, LarcheCahn1978}. The case involving multiple transport mechanisms \cite{CabrasEtAlJES2022} will be addressed in future work.

Denote with ${\cal P}_t \subset V$ an arbitrary region image of ${\cal P} \subset V_R$, and with 
$\partial {\cal P}_t$ and $\partial {\cal P}$ their respective boundaries.
{Let ${\phi} (\vect{x},t) \in {\cal P}_t$  and ${\phi}_R (\vect{X},t) \in {\cal P}$ 
be two generic tensorial functions of any order. If   
	\begin{equation}
		\int_{{\cal P}_t} {\phi} (\vect{x},t) \; {\rm d}v = 
		\int_{{\cal P}} {\phi}_R ( \vect{X}, t ) \; {\rm d}V \; ,
	\end{equation}
then the following transformation rule applies
	\begin{equation}
	\label{quantity_transformation_V}
		{\phi}_R (\vect{X},t)  =  J ( \vect{X}, t ) \, {\phi} ( \vect{x} ( \vect{X}, t ) ,t) \, .
	\end{equation}
}

\bigskip
Let $\vect{\varphi} (\vect{x},t)$ be a spatial vector, and $\vect{x}$ a point on an oriented surface $\partial{\cal P}_t$ 
defined by the direction of the outward normal vector $\vect{n}$. 
Denote with $\vect{n}_R$ the outward normal vector to the oriented surface $\partial{\cal P}$. 
The referential counterpart $\vect{\varphi}_R (\vect{X},t)$, with $\vect{X} \in \partial{\cal P}$, is defined by the identity
	\begin{equation}
	\label{vect_transformation_first}
		\int_{ \partial{\cal P}_t} \vect{\varphi} (\vect{x},t) \cdot \vect{n} \; {\rm d}a = 
		\int_{ \partial{\cal P}} \vect{\varphi}_R ( \vect{X}, t ) \cdot \vect{n}_R \; {\rm d}A \; ,
	\end{equation}
imposed on all $\partial{\cal P}_t \subset \partial{\cal B}_t$.
Due to the Nanson's formula for area changes, 
$\vect{n} \; {\rm d}a = J \, \Def^{\rm - T} \, \vect{n}_R  \; {\rm d}A $, Eq. \eqref{vect_transformation_first}
implies
	\begin{equation}
	\label{vect_transformation}
		\vect{\varphi}_R  (\vect{X},t) = 
		J ( \vect{X}, t ) \, \Def^{\rm - 1} ( \vect{X}, t ) \, \vect{\varphi} ( \vect{x} ( \vect{X}, t ) ,t)   \; .
	\end{equation}

\subsection{Reaction kinetics}
\label{sec:reaction kinetics}

Depict an arbitrary $i$-th chemical reaction in the form
\begin{equation}
\label{eq:reaction}
\sum_{\text{reactants}} |\nu_{ir}| A_r \rightleftharpoons \sum_{\text{products}} |\nu_{ip}| B_p ,
\end{equation}
where \(A\) and \(B\) label the different species and the \(\nu\)'s are the stoichiometric coefficients for reaction $i$. As in \cite{deGrootBook}, they are defined to be positive when species appear as a product in the reaction, negative when reactant.

\bigskip
Denote with $\zeta^\alpha_R$ the dimensionless ratio between referential quantities
\begin{equation}
\label{eq:zeta:1}
\zeta^\beta_R 
=
 \frac{ c^\beta_R  /  c^\beta_{sat,R}  }{ 1 - c^\beta_R  /  c^\beta_{sat,R} }
=
 \frac{ c^\beta_R   }{  c^\beta_{sat,R} - c^\beta_R  }
\; ,
\end{equation}
which, by virtue of Eq.~\eqref{eq:prop:cmax}, remains invariant under configuration changes
\begin{equation}
\label{eq:zeta:2}
\zeta^\beta_R = \zeta^\beta =
 \frac{ c^\beta   }{  c^\beta_{sat} - c^\beta }
\; .
\end{equation}
The law of mass action \cite{deGrootBook,GARCIACOLIN1985363} states that the kinetics of reaction in the reference configuration is well
modeled via the following law
\begin{equation}
\label{eq:mass_action_law}
w_R^{i} (\vect{x},t)  = {k^i_f}_R \prod_r {\zeta_R^r}^{|\nu_{ir}|} - {k^i_b}_R \prod_p {\zeta^p_R}^{|\nu_{ip}|} ,
\end{equation}
where $w_R^{i} (\vect{x},t) $ is the referential reaction rate at time \(t\), and \({k^i_f}_R\), \({k^i_b}_R\) are the so-called forward and backward rate constants of reaction $i$ in the reference configuration.
Equation \eqref{eq:mass_action_law} extends to a generic reaction the chemical kinetics for trapping that was detailed in \cite{SalvadoriEtAlJMPS2018,ARRICCA2023105425}.

The reaction rate $w_R^i$, being defined over a volume, transform according to 
Eq. \eqref{quantity_transformation_V}, 
	\begin{equation}
	\label{transformation_c_w_s}
		w_R^i (\vect{X},t) = J \, w^i \; .
	\end{equation}
Since $\zeta^\beta_R = \zeta^\beta$, eq. \eqref{transformation_c_w_s} reflects on the forward and backward rate constants as usual \cite{ARRICCA2023105425}, i.e., 
\begin{equation}
\label{eq:curr:mass_action_law}
w^{i} (\vect{x},t)  = {k^i_f} \prod_r {\zeta^r}^{|\nu_{ir}|} - {k^i_b} \prod_p {\zeta^p}^{|\nu_{ip}|} ,
\end{equation}
with
$$ 
 {k^i_f}_R = J \,  {k^i_f} \; ,
 \qquad
 {k^i_b}_R = J \,  {k^i_b} \; .
 $$

\bigskip
Energy storage in conversion-type electrodes offers a compelling setting for applying 
eq.~\eqref{eq:mass_action_law} to solid mixtures undergoing substantial volume changes, 
as described by eq.~\eqref{eq:def:Jc}, during chemical reactions. 
A representative example is the insertion of atomic sodium into tin, which gives rise 
to a sequence of intermediate compounds and the corresponding chemical reactions:
\begin{align}
&
 {\rm Na + 3 \, Sn}  \quad \rightleftarrows  \quad   {\rm NaSn_3}
\label{eq:IonizationReaction:b}
\\ &
 {\rm 8\, Na + NaSn_3 + Sn} \quad \rightleftarrows \quad  {\rm Na_{9}Sn_4}  
\label{eq:IonizationReaction:c}
\\ &
 {\rm 6\, Na + Na_9Sn_4} \quad \rightleftarrows \quad  {\rm Na_{15}Sn_4}
\; .  
\label{eq:IonizationReaction:d}
\end{align}
Experimental investigations \cite{WangEtAlNANOLETTERS2012}  show that the volumetric expansion is
about 60\% in the first step (eq. \eqref{eq:IonizationReaction:b}) and 420\% after reaction \eqref{eq:IonizationReaction:d}.
Atomic sodium can then diffuse much faster than any alloy, for its steric dimension. In fact, we may even assume that transport of alloys via solid-state diffusion is a secondary effect.

\bigskip
%
The distinction between the ``apparent'' and ``real'' saturation concentrations, introduced in 
eqs.~\eqref{eq:intermediate:saturation:2} and \eqref{eq:intermediate:saturation}, bears 
significant consequences for the definition of the proper form of the mass action law. As previously observed, the ratio
\[
\frac{c^\beta_R}{c^\beta_{sat,R}} = \frac{c^\beta}{c^\beta_{sat}}
\]
is influenced by the presence of other constituents into $c^\beta_{sat}$ through its dependence on $J^c$. This drawback 
can be circumvented by formulating the mass action law \eqref{eq:mass_action_law} in terms of the 
ratio
\[
\frac{c^\beta}{c^{\beta,{real}}_{sat}} 
= \frac{c^\beta_R}{ c^\beta_{sat,R} } \, \frac{ \omega^\beta }{ J^c }
\; ,
\]
thus adopting the factor
\begin{equation}
\label{eq:zeta:real}
\zeta^{\beta,{real}}
=
 \frac{ c^\beta  /  c^{\beta, {real}}_{sat}  }{ 1 - c^\beta  /  c^{\beta, {real}}_{sat}  }
=
 \frac{ {c^\beta_R} / { c^\beta_{sat,R} }  }{  \frac{ J^c }{ \omega^\beta } - {c^\beta_R} / { c^\beta_{sat,R} } \, }
=
 \frac{ \zeta^{\beta}_R  }{  \frac{ J^c }{ \omega^\beta } +( \frac{ J^c }{ \omega^\beta } -1)  \,  \zeta^{\beta}_R  }
\; ,
\end{equation}
rather than $\zeta^\beta_R=\zeta^\beta$, since $c^{\beta,{real}}_{sat}$ is an intrinsic property of 
species $\beta$ and remains unaffected by the composition of the surrounding mixture. Eq. \eqref{eq:zeta:real} can be inverted, 
providing
\begin{equation}
\label{eq:zeta:inv}
 \zeta^{\beta}_R   
 = 
\frac{ J^c }{ \omega^\beta } \;  \frac{ \zeta^{\beta,{real}}  }{  1+ \zeta^{\beta,{real}} - \frac{ J^c }{ \omega^\beta } \; \zeta^{\beta,{real}}   }
\; .
\end{equation}
to be inserted into eq. \eqref{eq:mass_action_law} prior to  the transformation rule \eqref{transformation_c_w_s}.

\subsection{Mass balance}
\label{sec:Mass balance}

Mass balance equations for the species can be formulated either in terms of densities or concentrations, within the reference configuration, for any subregion ${\cal P} \subset {V_R}$:
\begin{equation}
\label{eq:integral_mass_balance}
	\frac{{\rm d}}{{\rm d}t} \int_{\cal P}  c^\alpha_R (\vect{X}, t) \, {\rm d}V
	+ \int_{\partial{\cal P}} \massflux^\alpha_R(\vect{X}, t) \cdot \vect{n}_R \, {\rm d}A
	= \sum_i \int_{\cal P} \nu_{i\alpha} \, w_R^{i}(\vect{X}, t) \, {\rm d}V \; .
\end{equation}
The fluxes $\massflux_{\alpha_R}$ are defined relative to the unique advection velocity of the solid mixture.  
Chemical reactions act as sources or sinks of mass for individual species, without changing the total mass of the system, as expressed by the condition
\begin{equation}
\label{eq:mass_balance:chemical_reactions}
	\sum_\alpha \kappa_\alpha
 	\sum_i \nu_{i\alpha} \, w_R^{i}(\vect{X}, t) \, {\rm d}V = 0 \; ,
\end{equation}
where $\kappa_\alpha$ denotes the molar mass of species~$\alpha$. Condition \eqref{eq:def:ref_concentration_dependence} holds, as an alternative form of the total mass conservation --- see also \cite{Ganser2019} and the appendices therein.
After application of the divergence theorem, Eq. \eqref{eq:integral_mass_balance} takes the form
	\begin{equation}
	\label{integral_mass_balance_}
		\int_{ {\cal P}}  { \partial \over \partial t } \, c^\alpha_R ( \vect{X}, t ) \; {\rm d}V +
		\int_{ {\cal P}} \Divergence{ \massflux^\alpha_R ( \vect{X}, t) } \; {\rm d}V 
		=
 		\sum_i \int_{ {\cal P}}   \nu_{i\alpha}  \, w_R^{i} ( \vect{X}, t ) \; {\rm d}V 
		\; .
	\end{equation}
Since it holds for any region ${\cal P} \subset {V_R}$, it localizes as 
	\begin{equation}
	\label{massbalanceequations_ref}
		 \frac{ \partial }{ \partial t } \, c^\alpha_R ( \vect{X}, t ) +
		 \Divergence{ \massflux^\alpha_R ( \vect{X}, t) }  
		 =
 		\sum_i  \nu_{i\alpha}  \, w_R^{i} ( \vect{X}, t )  
		\; .
	\end{equation}
The mass flux $\massflux_{L_R}$ follows the rule stated by Eq. \eqref{vect_transformation}, namely,
	\begin{equation}
	\label{flux_transformation} 
		 \vect{h}_{\alpha_R} (\vect{X},t)  = J \, \Def^{-1} \,  \vect{h}_{\alpha}  \; .
	\end{equation}
%
%
The volume fractions of the phases can be derived from the concentrations of the constituent species, according to the relationships that link species to phases. In the representative case of tin sodiation, for example, five species can be identified ($\mathrm{Na}$, $\mathrm{Sn}$, $\mathrm{NaSn_3}$, $\mathrm{Na_9Sn_4}$, and $\mathrm{Na_{15}Sn_4}$), among which only one --- namely $\mathrm{Na}$ --- is capable of diffusing within the $\mathrm{Sn}$ phase. Consequently, four distinct phases may be defined: $\mathrm{NaSn_3}$, $\mathrm{Na_9Sn_4}$, $\mathrm{Na_{15}Sn_4}$, and the solid solution composed of $\mathrm{Sn}$ and the mobile $\mathrm{Na}$ species. 
The relationships between the concentrations of the various species and the corresponding phase volume fractions are given by Eq.~\eqref{eq:def:ref_volume_fraction}, subject to the constraint expressed in Eq.~\eqref{eq:prop:volume_fractions}.

\subsection{Referential balance of momentum}
\label{sec:Balance of momentum}

The referential local form the balance of linear momentum reads
\begin{equation}
\label{eq:ref_linearmomentum}
\Divergence{ \tensor{P}} +  {{\vect{B}}} = \totaltimed{   \rho_R \, \vect{v} \,    } \;
,
\qquad
\vect{X} \in  V_R
\end{equation}
where $\vect{B}$ denotes the external body forces per unit reference volume in the material description, $\vect{v}$ the referential advection velocity, and $\rho_R$ the referential density of the system. Since the restriction in Eq.~\eqref{eq:mass_balance:chemical_reactions} holds, no external mass supply is introduced at any point $\vect{X} \in V_R$. Moreover, because the referential molar volumes of all species are assumed to be arbitrary and equal, that is, ${\Omega}^\alpha_R = {\overline{\Omega}}_R$, the referential density $ \rho_R$ remains unchanged. This invariance does not extend to the actual density in the current configuration, which varies due to the distinct real molar volumes associated with each phase.

${ \tensor{P}} $ is the nominal stress tensor (first Piola-Kirchhoff stress tensor),  
obeying the symmetry condition 
\begin{equation}
\label{eq:ref_angularmomentum}
\tensor{P} \tensor{F}^T =  \tensor{F} \tensor{P}^T \; ,
\end{equation} 
%
which is the balance of angular momentum. 

\section{Thermodynamics}
\label{sec:thermo}

A comprehensive thermodynamic analysis of transport, reaction, and mechanical coupling has been conducted for neutral species in \cite{ARRICCA2023105425} and for charged species in \cite{serpelloniEtAl2025EJMA}. 
As no substantial developments beyond those works are introduced here, the reader is referred to those studies for detailed derivations. 
In the present note, we confine ourselves to summarizing the most pertinent results.

\subsection{Energy balance.}
\label{subsec:firstprinciple}

%
By defining $u_R$ as the specific internal energy (per unit reference volume), we write
the local form of the energy balance as 
	\begin{equation}
	\begin{aligned}
	\label{energe_balance_ref_loc}
		{ {\rm d} {u}_R \over {\rm d} t }  &= 
		\firstpiola : \dot{\Def} + {\heatsource}_R -  \Divergence{\heatflux_R} 
		+
		\sum_\alpha
		\uchempot_\alpha \,   \frac{ \partial c^\alpha_R }{ \partial t } 
		- 
		\sum_\alpha
		\massflux^\alpha_R \cdot \Gradient{\uchempot_\alpha} 
		-
		 \sum_{i,\alpha}  \nu_{i\alpha} \, \uchempot_\alpha\, w_R^{i}(\vect{X}, t) 
		 \; .
	\end{aligned}
	\end{equation}
The scalar ${\uchempot}_\alpha$ is the energy provided by a unit supply 
of moles of species $\alpha $, whereas ${\heatsource}_R$ and $\heatflux_R$ represent the heat supplied by external agencies and the heat flux vector, 
respectively.

\subsection{Entropy imbalance.}
\label{subsec:secondprinciple}

Similarly to the energy balance, we denote with $\entropy_R$ the specific net internal entropy (per unit reference volume).
The local form of the second law of thermodynamics can be stated as
	\begin{equation}
	\begin{aligned}
	\label{local_entropy_imbalance}
		T \, { {\rm d} {\entropy}_R \over {\rm d} t } & - \heatsource{_R} + \Divergence{\heatflux_R} - 
		{1 \over T} \, \heatflux_R \cdot \Gradient{T} 
		- 
		T \, 
		\sum_\alpha
		\etachempot_\alpha \,   \frac{ \partial c^\alpha_R }{ \partial t } 
		+
		T \,
		\sum_\alpha
		\massflux^\alpha_R \cdot \Gradient{\etachempot_\alpha} 
		\\&
		- 
		T 
		 \sum_{i,\alpha}  \nu_{i\alpha} \, \etachempot_\alpha\, w_R^{i}(\vect{X}, t)  
		 \ge 0
		 \; .
	\end{aligned}
	\end{equation}
Taking advantage of Eq. \eqref{energe_balance_ref_loc} to replace the factor 
$- \heatsource{_R} + \Divergence{\heatflux_R} $ in Eq. \eqref{local_entropy_imbalance}, we can rephrase the 
second law of thermodynamics as
	\begin{equation}
	\label{CD_inequality_step_x}
	\begin{aligned}
		T \, { {\rm d} {\entropy}_R \over {\rm d} t } 
		&-
		{ {\rm d} {u}_R \over {\rm d} t } 
		+
		\firstpiola : \dot{ \Def} 
		+ 
		\sum_\alpha
		( \uchempot_\alpha - T \, \etachempot_\alpha)  \,   \frac{ \partial c^\alpha_R }{ \partial t } 
		-
		\sum_\alpha
		\massflux^\alpha_R \cdot \Gradient{\uchempot_\alpha} 
		\\&- {1 \over T} \, \heatflux_R \cdot \Gradient{T} 
		+
		T \,
		\sum_\alpha
		\massflux^\alpha_R \cdot \Gradient{\etachempot_\alpha} 
		+
		 \sum_{i,\alpha}  \nu_{i\alpha} \, ( \uchempot_\alpha -T \,  \etachempot_\alpha ) \, w_R^{i}(\vect{X}, t)  
		\ge 0
		\; .
	\end{aligned}
	\end{equation}
By denoting 
	\begin{equation}
	\label{chempotential}
		\mu_\alpha = \uchempot_\alpha - T \, \etachempot_\alpha \; ,
	\end{equation}
	\begin{equation}
	\label{affinity}
 		A^i  =   \sum_{\alpha}  \nu_{i\alpha} \, \mu_\alpha  \; ,
	\end{equation}
we can cast the entropy imbalance \eqref{CD_inequality_step_x} in the form
	\begin{equation}
	\label{CD_inequality_step_xy}
	\begin{aligned}
		T \, { {\rm d} {\entropy}_R \over {\rm d} t } 
		&-
		{ {\rm d} {u}_R \over {\rm d} t } 
		+
		\firstpiola : \dot{ \Def} 
		+ 
		\sum_\alpha
		\mu_\alpha \,   \frac{ \partial c^\alpha_R }{ \partial t } 
		-
		\sum_\alpha
		\massflux^\alpha_R \cdot \Gradient{\uchempot_\alpha} 
		\\&- {1 \over T} \, \heatflux_R \cdot \Gradient{T} 
		+
		T \,
		\sum_\alpha
		\massflux^\alpha_R \cdot \Gradient{\etachempot_\alpha} 
		+
		 \sum_{i}  A^i \, w_R^{i}(\vect{X}, t)  
		\ge 0
		\; .
	\end{aligned}
	\end{equation}
Following \cite{SalvadoriEtAlJMPS2018}, we introduce a new referential heat flux as
	\begin{equation}
	\label{newheatflux}
		\newheatfluxR = \heatflux_R + T \, \sum_\alpha \etachempot_\alpha \, \massflux^\alpha_R \; ,
	\end{equation}
which leads to the alternative localized referential form of the entropy imbalance
	\begin{equation}
	\label{entropy_imbalance_ref_loc_new}
	\begin{aligned}
		T \, { {\rm d} {\entropy}_R \over {\rm d} t } 
		&-
		{ {\rm d} {u}_R \over {\rm d} t } 
		+
		\firstpiola : \dot{ \Def} 
		- 
		{1 \over T} \, \newheatfluxR \cdot \Gradient{T} 
		+ 
		\sum_\alpha
		\mu_\alpha \,   \frac{ \partial c^\alpha_R }{ \partial t } 
		-
		\sum_\alpha
		\massflux^\alpha_R \cdot \Gradient{\mu_\alpha} 
		+
		 \sum_{i}  A^i \, w_R^{i}(\vect{X}, t)  
		\ge 0
		\; .
	\end{aligned}
	\end{equation}
%

\subsection{Clausius-Duhem inequality.}
\label{CD_inequality}

From the definition of the Helmholtz free energy density (per unit referential volume), 
	\begin{equation}
		\label{psi_R}
		\helmholtz_R = u_R - T \, \entropy_R \; ,
	\end{equation}
the first two terms of the localized referential form of the entropy imbalance \eqref{entropy_imbalance_ref_loc_new} 
can be re-written as
	\begin{equation}
		T \, { {\rm d} {\entropy}_R \over {\rm d} t }  - { {\rm d} {u}_R \over {\rm d} t } = - 
		{ {\rm d} {\helmholtz}_R \over {\rm d} t }  - { \partial {T} \over \partial t } \, \entropy_R \; .
	\end{equation}

We select the following state variables: temperature $T$, concentrations
of species $c^\alpha_R$, chemo-thermo-elastic right Cauchy-Green strain tensor $\rightCG$. For the sake of simplicity we limit our investigation to elastic materials, 
interested readers may refer to \cite{ARRICCA2023105425} for more general formulations of inelastic 
constitutive laws \cite{GurtinFriedAnand, TadmoreBook1, paolucci2016, HolzapfelBook, SIMOCMAME1988a, 
SIMOCMAME1988b}.
Accordingly,
	\begin{equation}
	\label{psi_internal_variables}
		\helmholtz_R = \helmholtz_R \left( T, c^\alpha_R, \rightCG \right) 
	\end{equation}
and the time derivative of $\helmholtz_R$ writes as 
	\begin{equation}
	\label{time_derivative_psi_funct_dep}
		{ {\rm d} {\helmholtz}_R \over {\rm d} t } = 
		{\partial \helmholtz_R \over \partial \rightCG} : { \partial { \rightCG } \over \partial t}  
		+
		\sum_\alpha
		{\partial \helmholtz_R \over \partial c^\alpha_R} \, { \partial c^\alpha_R \over \partial t }
		+
		{\partial \helmholtz_R \over \partial T} \,   { \partial T \over \partial t } 
		\; .
	\end{equation}
Making use of the second Piola-Kirchhoff stress $\secondpiola = \Def^{-1} \, \firstpiola$, eqs \eqref{entropy_imbalance_ref_loc_new} and
\eqref{time_derivative_psi_funct_dep} can be combined
in the Clausius-Duhem inequality in the form
	\begin{equation}
	\begin{aligned}
	\label{CD_inequality_2}
		\left(  {1 \over 2} \, \secondpiola - {\partial \helmholtz_R \over \partial \rightCG}  \right) 
		&: { \partial { \rightCG } \over \partial t}  
		+
		\sum_\alpha
		\left( \mu_\alpha - {\partial \helmholtz_R \over \partial c^\alpha_R}  \right) \,  { \partial c^\alpha_R \over \partial t }
		-
		\left( \entropy_R +  {\partial \helmholtz_R \over \partial T}  \right) \,{ \partial T \over \partial t } 
		\\&
		- 
		{1 \over T} \, \newheatfluxR \cdot \Gradient{T} 
		- 
		\sum_\alpha \massflux^\alpha_R \cdot  \Gradient{\mu_\alpha} 
		- 
		 \sum_{i}  A^i \, w_R^{i}(\vect{X}, t)  
		\ge 0 \; .
	\end{aligned}
	\end{equation}
This inequality must hold for any value of the time derivative of the strain tensor $\rightCG$, of the concentrations 
$c^\alpha_R$, and of the temperature $T$. 
The following thermodynamic restrictions thus arise
	\begin{subequations}
	\label{thermo_restrictions}
	\begin{align}
	\label{thermorestrictions_1}
		\secondpiola &=   2 \,  { \partial \helmholtz_R \over  \partial \rightCG } \; , 
		\\
	\label{thermorestrictions_1_1}
		\entropy_R &= - {\partial \helmholtz_R \over \partial T}  \; ,
		\\
	\label{thermorestrictions_2} 
		\mu_\alpha &=   {\partial \helmholtz_R \over \partial c^\alpha_R}  \; .
	\end{align}
	\end{subequations}
What remains of the Clausius-Duhem inequality \eqref{CD_inequality_2} establishes irreversible processes.
As we cannot place any restriction on the values of ${\rightCG }$, $c^\alpha_R$, and $T$,
the same applies for $\Gradient{\mu_\alpha}$, $\Gradient{T}$, and 
$w_R^{i}$.
As a consequence, restrictions on $\massflux^\alpha_R$, $\newheatfluxR$, 
and $A^i$ are determined.
The internal entropy production (multiplied by $T$), can be written with the usual dissipative 
structure \cite{PrigogineNL1977}, namely
	\begin{equation}
	\label{TDinequalities}
		-
		\underbrace{
		\sum_\alpha
		\massflux^\alpha_R \cdot  \Gradient{ \mu_\alpha } 
		}_{\mbox{{diffusive}}} - 
		\underbrace{
		{1 \over T} \,  \newheatfluxR   \cdot \Gradient {T}
		}_{\mbox{thermal}} - 
   		 \underbrace{   
		  \sum_{i}  A^i \, w_R^{i}(\vect{X}, t)  
		 }_{\mbox{chemical}}
    		\ge 0 \; .
	\end{equation}
No coupling between fluxes and thermodynamic forces of different tensorial order is imposed in view of the 
assumption of Curie's symmetry principles \cite{deGrootBook}. 
Therefore the the following three conditions arise from the internal entropy production \eqref{TDinequalities}
	\begin{subequations}
	\begin{align}
	\label{thermorestrictions_vect}
		\sum_\alpha
		\massflux^\alpha_R \cdot  \Gradient{ \mu_\alpha } 
		+ 
		{1 \over T} \, \newheatfluxR  \cdot \Gradient {T} &\le 0 \; ,   
		\\
	\label{thermorestrictions_scal}
		 \sum_{i}  A^i \, w_R^{i}(\vect{X}, t)  & \le 0 \; . 
	\end{align}
	\end{subequations}
In view of thermodynamic restrictions \eqref{thermorestrictions_2}, the scalars 
$\mu_{\alpha}$ declared in Eq. \eqref{chempotential} are seen to be chemical potentials,
whereas $A^i$ is the affinity of the $i$-th chemical reaction. Specifications for ${\uchempot}_\alpha $ and ${\etachempot}_\alpha $ can be given as in \cite{ARRICCA2023105425}.

\subsection{Strain Decomposition.}
\label{strain_decomposition}

The multiplicative decomposition \eqref{eq:multiplicative_decomposition} was introduced in section \ref{subsec:intermediate} to elaborate the swelling factor $J^c$ in eq. \eqref{eq:def:Jc}. Limiting our focus on elastic materials, the chemo-thermo-elastic deformation gradient $\Def$ can be subject to the multiplicative 
decomposition
	\begin{equation}
		\label{F^cte}
		\Def = \Def^e  \, \Def^{th} \, \Def^c \; ,
	\end{equation}
with $\Def^e$ the local reversible elastic distorsion, $\Def^{th}$ the thermal contribution to the local distorsion, and $\Def^c$ has been already defined. The non-elastic contributions are henceforth assumed to have a spherical form, hence
	\begin{subequations}
	\label{F_s_th}
	\begin{align}
	\label{F^s}
		\Def^c = \lambda_c \, \Id  \; ,
		\\
	\label{F^th}
		\Def^{th} = \lambda_{th} \, \Id \; ,
	\end{align}
	\end{subequations}
with $\lambda_c$ and $\lambda_{th}$ positive coefficients termed chemical swelling and thermal 
stretch, respectively. In view of eq. \eqref{eq:def:Jc}, the swelling volumetric Jacobian and its evolution in time write as
	\begin{subequations}
	\begin{align}
	\label{Js}
		& \lambda_c^3 = J^{c}  =  \sum_\alpha  c_R^\alpha \;  \Omega^\alpha  \; , 
		\\ 
	\label{dot_Js}
		& \dot{J}^c =  \sum_\alpha  \frac{\partial c_R^\alpha}{\partial t} \;  \Omega^\alpha \; .
	\end{align}
	\end{subequations}
By defining with $f^{th} (T)$, the function which identifies the thermal volumetric 
Jacobian,
	\begin{equation}
	\label{f_th}
		J^{th} = f^{th} (T)  \; ,
	\end{equation}
its time derivative writes as
	\begin{equation}
	\label{dot_J_th}
		\dot{J}^{th} = \gamma \, { \partial T \over \partial t }  
		\; ,
	\end{equation}
where $\gamma$ denotes the {thermal expansion coefficient},
	\begin{equation}
	\label{thermal_exp} 
		\thermalexp = {\partial f^{th} \over \partial T} \; .
	\end{equation}
A classical choice for the function $f^{th}(T)$ is given by the relation
	\begin{equation}
		f^{th} (T) = 1 + \thermalexp \, \left( T - T_0 \right) \;, 
	\end{equation}
with $T_0$ the reference (datum) temperature. 

\bigskip
The elastic part of $\rightCG$ can be made explicit. In view of the multiplicative decomposition of $\Def$ given by 
Eq. \eqref{F^cte}, and the definitions of the swelling and thermal deformation gradients,  $\Def^{c}$ and  $\Def^{th}$, 
shown in Eqs \eqref{F_s_th}. 
The chemo-thermo-elastic deformation gradient can be therefore re-written as
	\begin{equation}
		\Def 
		= \Def^e \, \lambda_{th} \, \lambda_c \; .
	\end{equation}
It follows that, the elastic right Cauchy-Green strain tensor takes the form
	\begin{equation}
	\label{C_e_def_in_progr}
		\rightCG^e = \Def^{e^{\rm T}}  \Def^e = 
		J^{c^{- {2 \over 3} } } J^{ {th}^{- {2 \over 3} } } \, \rightCG \; .
	\end{equation}

\subsection{Constitutive theory}
\label{const_theory}

Since the referential domain $V_R$ under consideration is composed of multiple subdomains, each characterized by its own volume fraction, 
the referential Helmholtz free energy density $\helmholtz_R$ can be expressed as the weighted average of the free energies of the individual solid phases $\beta$:
\begin{equation}
\label{eq:Helmholtz:average}
\helmholtz_R (T, c^\alpha_R, \rightCG)
= 
\sum_\beta \Phi^\beta \, \helmholtz^\beta_R (T, c^\alpha_R, \rightCG)
\; .
\end{equation}
As previously emphasized, we maintain a clear distinction between phases ($\beta$) and species ($\alpha$): 
the latter may diffuse within each phase and, in principle, influence its free energy $\helmholtz^\beta_R$. Besides the sodiation of tin anodes, mechanobiology provides several examples of multiphase solid systems in which proteins, growth-factors, and other species are transported and react.

\bigskip
As discussed in detail in \cite{SalvadoriEtAlJMPS2018,ARRICCA2023105425}, the constitutive formulation is founded on an additive decomposition 
of the Helmholtz free energy $\helmholtz^\beta_R$ at a material point $\vect{X}$:
\begin{equation}
\label{psi_splittin}
\helmholtz^\beta_R \left( T, c^\alpha_R, \rightCG \right) 
= \psi_0 
+ {\helmholtz^\beta_R}^{\, th} \left( T, c^\alpha_R \right)
+ {\helmholtz^\beta_R}^{\, diff} \left( T, c^\alpha_R, \rightCG \right)
+ {\helmholtz^\beta_R}^{\, el} \left( T, c^\alpha_R, \rightCG \right)
\; ,
\end{equation}
where $\psi_0$ denotes a reference value, 
${\helmholtz^\beta_R}^{\, th}$ represents the thermal contribution, 
${\helmholtz^\beta_R}^{\, diff}$ the diffusive contribution, 
and ${\helmholtz^\beta_R}^{\, el}$ the elastic contribution. 
Inelastic effects are disregarded here for simplicity. 
Since the present work introduces no further developments beyond \cite{SalvadoriEtAlJMPS2018,ARRICCA2023105425}, 
the reader is referred to those studies for a comprehensive discussion of the individual contributions. 
We only remark here that the definition of the dimensionless ratios $\zeta^\alpha$ and $\zeta^{\beta,{real}}$, discussed in 
equations \eqref{eq:zeta:1} and \eqref{eq:zeta:real} 
plays a crucial role in the formulation of the entropy of mixing as derived from statistical mechanics. 

\bigskip
As done in  \cite{ARRICCA2023105425}, to comply with the restriction \eqref{thermorestrictions_vect} 
we relate the ordinary heat flux $\heatflux_R$ to the temperature gradient via Fourier's law, and the mass fluxes $\massflux^\alpha_R$ to the related gradient $ \Gradient{ \mu_\alpha } $. Therefore
	\begin{subequations}
	\begin{align}
	\label{Ficks_law}
		\massflux^\alpha_R &= - \tensor{M}^\alpha_R(c^\alpha_R) \, 
		\left( \Gradient{ {\mu}_{\alpha} } + \etachempot_\alpha \, \Gradient{T} \right) \; ,
		\\
	\label{Fouriers_law}
		\heatflux_R &= -  \myK  \Gradient{ T }  \; .
	\end{align}
	\end{subequations}

To account for saturation and to satisfy the physical requirement that both the pure and the saturated phases lead to vanishing mobility, the mobility tensor $\tensor{M}^\alpha_R(c^\alpha_R) $
is chosen in the following isotropic nonlinear form  \cite{AnandJMPS2012} 
	\begin{equation}
	\label{mobility_tensor}
		\tensor{M}^\alpha_R(c^\alpha_R) = 
		\mobility_{\alpha} \, c^\alpha_R \, \left( 1 -  \Phi^\alpha_R \right) \Id \; ,
	\end{equation}
with $\mobility_{L} > 0$ the mobility of interstitial chemical species.
%

Furthermore, following what is proposed in \cite{ARRICCA2023105425}, the referential rate factor which describes the reactions are taken as
	\begin{subequations}
	\label{newmassactionlaw_k_LR_k_TR}
	\begin{align}
	\label{eq:newmassactionlaw_k_LR}
		 {k^i_b}_R &=  
 		 {\tilde  {k^i_b}}_R
		\; \prod_\alpha  
 		\, \exp \left( { 1 \over R \, T} \, {\partial  \psi^{\, el}_R  \over \partial c^\alpha_R } \right) 
 		\, \exp \left(-  {  \etachempot_T^0 \,  ( T - T_0 )  \over R \, T } \right)   
 		\, \exp \left( - {  c_{v \, T}^0  (T - T_0)^2  \over 2 \,R \, T \, T_0} \right)  
		  \; ,
 		\\ 
	\label{eq:newmassactionlaw_k_TR}
		 {k^i_f}_R &= 		  
 		 {\tilde  {k^i_f}}_R
		\; \prod_\alpha  
 		\, \exp \left( { 1 \over R \, T} \, {\partial  \psi^{\, el}_R  \over \partial c^\alpha_R } \right) 
 		\, \exp \left(-  {  \etachempot_T^0 \,  ( T - T_0 )  \over R \, T } \right)   
 		\, \exp \left( - {  c_{v \, T}^0  (T - T_0)^2  \over 2 \,R \, T \, T_0} \right)  
       		\; .
	\end{align}
	\end{subequations}
{This formulation is consistent with the usual mass action law described by the {\it van't Hoff} relation 
\cite{ShellBook2015}, which is recovered when elastic and swelling contributions vanish.}

\section{Governing equations}
\label{governing_equations}

The  governing equations are obtained by incorporating 
the constitutive prescriptions of the chemo-thermo-elastic second Piola-Kirchhoff stress tensor $\secondpiola$, 
the referential heat and mass flux vectors, $\heatflux_R$ and $\massflux^\alpha_R$, 
and the mass action law \eqref{eq:mass_action_law} with specifications \eqref{newmassactionlaw_k_LR_k_TR}, 
into the balance equations \eqref{massbalanceequations_ref}, \eqref{eq:ref_linearmomentum}, 
and \eqref{energe_balance_ref_loc}.
Governing equations are written in term of the state variables $T$, $c^\alpha_R$, and $\rightCG$. Neglecting inertia terms in eq.  \eqref{eq:ref_linearmomentum}, they read
	\begin{subequations}
	\label{GoverningEQS}
	\begin{align}
	    \label{governing_mass_balance}
	    &
		 \frac{ \partial  c^\alpha_R}{ \partial t }  \, +
		 \Divergence{ \massflux^\alpha_R \left( T, c^\gamma_R, \rightCG \right)  }  
		 =
 		\sum_i  \nu_{i\alpha}  \, w_R^{i} \left( T, c^\gamma_R, \rightCG \right) 
		 \; ,
	\\ &
	\label{governing_linear_mom_balance}
		\Divergence{\firstpiola \, \left( T, c^\gamma_R, \rightCG \right) } +  {{\vect{B}}} = \vect{0}  \;  \; ,
	\\ & \nonumber
		- T \, {\partial^2 \helmholtz_R \over \partial T^2} {\partial T \over \partial t } - \Divergence{\tensor{ \myK } 
		\Gradient{ T }} 
		=
		\\ & \qquad \qquad 
		T \, {\partial^2 \helmholtz_R \over \partial T \partial \rightCG } :  {\partial { \rightCG } \over \partial t}  
		+ 
		{\heatsource}_R  
		- 
		\sum_\alpha
		\massflux^\alpha_R \left( T, c^\gamma_R, \rightCG \right)  \cdot \Gradient{\uchempot_\alpha} 
		-
		 \sum_{i,\alpha}  \nu_{i\alpha} \, \uchempot_\alpha\, w_R^{i}(\vect{X}, t) 
	\label{generalized_heat_eq}
	\end{align}
	\end{subequations}
The boundary conditions 
	\begin{subequations}
    \label{BoundaryCnds}
	\begin{align}
		\massflux^\alpha_R \cdot \vect{n}_R &= \overline{h}^\alpha_{R} \qquad \vect{X} \in \partial^N V \; ,
		\\
    \label{BoundaryCndHeat}
		\heatflux_R \cdot \vect{n}_R &= \overline{ q }_R \qquad  \vect{X} \in \partial^N V \; ,
		\\
		\firstpiola \, \vect{n}_R &= 
		\overline{ \Traction }   \qquad  \vect{X} \in \partial^N V \; ,
	\end{align}
	\end{subequations}
are imposed along Neumann boundaries $ \partial^N V$. To ensure solvability of the problem, 
Dirichlet boundary conditions have to be enforced along the complementary boundary $ \partial^D V$, hence
	\begin{subequations}
    \label{InitialCnds} 
	\begin{align}
    \label{InCondTemp}
		T &= \overline{T} \qquad \vect{X} \in \partial^D V \; ,
		\\
    \label{InCondDisp}
		\vect{u} &= \overline{ \vect{u} } \qquad \vect{X} \in \partial^D V \; .
	\end{align}
	\end{subequations}

Initial conditions are usually imposed for the concentration of species, $c^\alpha_R (\vect{X}, t = 0)$, 
as well as temperature, $T (\vect{X}, t = 0)$. To comply 
with equilibrium thermodynamics these conditions are chosen to be uniform in the material and to ensure
 equilibrium with external species. Balance of momentum, together with boundary conditions, provide 
the necessary and sufficient equations and conditions to solve for $\vect{u}$ at $t = 0$.

\section{Concluding remarks}
\label{sec:conclisions}

The present note focuses on reactive solid mixtures, assuming that no relative motion occurs among the coexisting solid phases \cite{ateshian2007theory}. In this regard,
this note differs from unconstrained mixtures, in which each constituent may move independently of others \cite{truesdell1957sulle,truesdell1957sulleII,TruesdellToupin1960}, as well as from 
constrained mixtures where different constituents move together only in the current configuration \cite{humphrey2002constrained,Humphrey:2021aa}. We have formulated a finite strains, chemo-mechanical continuum theory that 
assumes that a single motion characterizes the kinematics of the body, whence a single reference configuration for can be defined. As shown in section \ref{subsec:LC}, the classical network 
model --- a chemo-mechanical framework established by Larch{\'e} and Cahn for equilibrium \cite{LarcheCahn1978,LarcheCahn1973} --- and its subsequent extensions addressing transport and chemical reactions \cite{ChesterEtAlIJSS2015,ChesterAnandJMPS2010,AnandJMPS2012,Drozdov2014b,AFSHAR2021104368,SalvadoriEtAlJMPS2018,ARRICCA2023105425} can be seen as a special case of the proposed
framework.

There is indeed a wide range of scientific and engineering applications in which tightly coupled physico-chemical processes --- mass transport,  mechanical deformation, chemical reactions, thermal decomposition, phase transformation, and gas-generation reactions --- govern performance in solid mixtures. 
A classical field of application is {\it{metallurgy}}, where phase transformations driven by diffusing 
species govern the development of microstructure and mechanical properties. Carbon diffusion in 
steels during carburizing or quenching leads to the formation of martensite and associated 
residual stresses, profoundly enhancing surface hardness and fatigue resistance 
\cite{CAVALIERE200926,Woowiec-KoreckaAMSE2023,YangNIT2012}. Similar chemo-mechanical coupling 
phenomena arise in nitriding processes, where nitrogen diffusion and nitride precipitation shape 
the performance of treated components.
{\it{Electrochemical energy storage}} is another major domain. In intercalation electrodes, the insertion and extraction of ions induce significant volumetric 
changes and stress fields \cite{MENDOZA20161,LUO2016759}. Even more 
pronounced are conversion-type electrodes, where the host material undergoes extensive chemical 
transformations and structural rearrangements, often accompanied by several hundred percent 
volume expansion \cite{AFSHAR2021104368,WangEtAlNANOLETTERS2012}. 
Reactive mixtures also play a role in {\it{geomechanics}}, where mineral dissolution and precipitation 
alter porosity and mechanical stability in rocks, with implications for subsurface storage, 
hydrothermal processes, and induced seismicity \cite{STEEFEL2005539}. In {\it{mechanobiology}},
chemo-mechanical models of tissues --- such as cartilage undergoing remodeling driven by biochemical 
signals and mechanical loading --- can be interpreted as reactive solid mixtures whose internal 
constituents evolve via synthesis and degradation reactions 
\cite{RAHMAN2023107419,Pierce:2016aa}.
{\it{Solid rocket propellants}} is another prominent class of reactive solid mixtures. Composite propellants typically consist of a polymeric binder matrix (e.g., HTPB) embedding crystalline oxidizers such as ammonium perchlorate (AP) and metallic fuels (e.g., aluminum powder). Upon ignition, heterogeneous reactions occur at the solid–gas interface: oxidizer decomposition, combustion of metal particles, and binder pyrolysis form a multiphase reactive system whose regression rate depends sensitively on local temperature, species diffusion, and microstructural distribution. The interplay between reaction kinetics, heat transfer, and evolving porosity links the macroscopic burn characteristics to the underlying chemistry and mechanics. These systems have been extensively studied for aerospace propulsion, ballistic applications, and next-generation high-energy materials \cite{SuttonRocketPropulsion, KubotaPropellants,DENNIS2019115646}.

These diverse applications underscore the broad relevance of reactive solid mixtures and 
motivate the development of rigorous theoretical frameworks capable of capturing the 
interplay among reaction kinetics, transport phenomena, and large deformations. The 
fundamental balance laws have been established in Section~\ref{sec:Balancelaws}, while the 
thermodynamic treatment of transport, reaction, and mechanochemical coupling presented in 
Section~\ref{sec:thermo} builds upon the comprehensive analyses carried out for neutral 
species in~\cite{ARRICCA2023105425} and for charged species in~\cite{serpelloniEtAl2025EJMA}. 
A central contribution of this work is the observation that phase volumes remain unchanged in 
the reference configuration, even as the referential concentrations of reactive species vary. 
This convention preserves a stress-free reference state throughout chemical transformations, 
yet it profoundly influences the kinematics of swelling and the formulation of intermediate 
configurations. Indeed, the implications of defining $J^c$ as in 
eq.~\eqref{eq:def:Jc} constitute the principal outcome of this study, extending classical 
expressions adopted in several prior works (e.g., 
\cite{ChesterAnandJMPS2010,Drozdov2014b,DiLeoAnand2013,ARRICCA2023105425,Ganser2019,AnandJMPS2012}).


The concepts developed herein will be further applied to the predictive modeling of cellular 
motility --- using the reactive mixture framework introduced in 
\cite{SalvadoriSR2024,SerpelloniEtAlAMS2021,BONANNO2023105273} --- and to the description of fibrin 
assembly and contraction during blood clot formation 
\cite{Kim:2017aa,Kliuchnikov:2025aa,AlberJRInt2008}. 
These biological networks transmit mechanical stresses either through active contraction 
\cite{Jallon:2025aa} (as observed in fibrin 
\cite{FILLA2024105750,GU2025105998}) or through compression mechanisms, such as those occurring 
in the lamellum between the cell membrane and focal adhesions 
\cite{PollardARBB2000,KawskaPNAS2012,Blanchoin:2014uf}. 
A defining feature of such networks is their dynamic nature: they are assembled chemically 
when required and dismantled once their mechanical role has been fulfilled. 
While the chemo-mechanical framework proposed in this work can describe the formation of new 
species endowed with their own molar volume, it stands in marked contrast with the behavior of 
biological cytoskeletal or fibrin networks. In these systems, the solid phase responsible 
for transmitting and redistributing mechanical loads cannot be specified a priori, nor can it 
be assumed to remain fixed in a prescribed reference configuration.

\bigskip \noindent
\textbf{ \large Acknowledgements}

\bigskip
This research is funded by: the European Union’s Horizon 2020 research and innovation programme HORIZON-EUROHPC project “dealii-X: an Exascale Framework for Digital Twins of the Human Body” (topic HORIZON-EUROHPC-JU-2023-COE-03-01). Project Grant 101172493, the Mechanobiology research center at UNIBS (https:// www. mechanobiology- unibs. it/) through the support of companies COMSOL, Copan, and Antares Vision, the Ferriera Valsabbia foundation and Comipont through generous donations to fund studies in the field of Mechanobiology. The work has been carried out as part of INdAM (Istituto Nazionale di Alta Matematica) GNFM activities.
AS gratefully acknowledges the collaboration of the student G. Luciani for a bibliographic search on the modeling of cartilage.

\bibliographystyle{unsrt}
\bibliography{/Users/albertosalvadori/Bibliography/Bibliography}

\end{document}